\documentclass[prl,showkeys,showpacs,groupedaddress,twocolumn]{revtex4}
\usepackage{amssymb}
\usepackage{graphicx}
\usepackage{dcolumn}
\usepackage{bm}
\usepackage{amsmath}
\usepackage{subfigure}
\usepackage{float}
\usepackage{color}
\begin{document}

\title{Atom-interferometric measurement of Stark level splittings}

\author{Limei Wang}
\author{Hao Zhang}
\author{Linjie Zhang}
\author{Georg Raithel$^{1}$}
\author{Jianming Zhao}
\thanks{Corresponding author: zhaojm@sxu.edu.cn}
\author{Suotang Jia}
\affiliation{State Key Laboratory of Quantum Optics and Quantum Optics Devices, Institute of Laser spectroscopy, Shanxi University, Taiyuan 030006, P. R. China}
\affiliation{$^{1}$Department of Physics, University of Michigan, Ann Arbor, Michigan 48109-1120, USA}
\date{\today}

\begin{abstract}
Multiple adiabatic/diabatic passages through avoided crossings in the Stark map of cesium Rydberg atoms are employed as beam splitters and recombiners in an atom-interferometric measurement of energy-level splittings. We subject cold cesium atoms to laser-excitation, electric-field and detection sequences that constitute an (internal-state) atom interferometer.
For the read-out of the interferometer we utilize state-dependent collisions, which selectively
remove atoms of one kind from the detected signal.
We investigate the dependence of the interferometric signal on timing and field parameters, and find good agreement
with time-dependent quantum simulations of the interferometer. Fourier analysis of the interferometric signals yield coherence frequencies that agree with corresponding energy-level differences in calculated Stark maps. The method enables spectroscopy of states that are inaccessible to direct laser-spectroscopic observation, due to selection rules, and has applications in field metrology.

\end{abstract}


\maketitle

Matter-wave interference offers exquisite sensitivity to measure fields and atomic or molecular interactions. Examples based on atom inertia include
atom-interferometric gravimetry~\cite{angelis,hughes,Geiger}, gradiometry~\cite{McGruirk,Kohel}, and Sagnac-type rotation sensors (gyroscopes)~\cite{Lenef1997,Gustavson}. Such devices usually involve laser-based beam splitters to coherently split and recombine wavefunctions. Another well-known example of matter-wave interference is the superconducting quantum interference device (SQUID)~\cite{Fagaly}, which engages  the vector potential to measure magnetic-field-induced phases in matter waves~\cite{rkVarma,RK,RKV}. Applications of Ramsey's separated-oscillatory-field method~\cite{Ramsey}, which employs quantum interference between field-coupled internal states of a quantum particle, are abound in spectroscopy and time metrology~\cite{ADCronin,KHornberger}.

\begin{figure}[htb]
\vspace{-0.1em}
\centering
\includegraphics[width=0.5\textwidth]{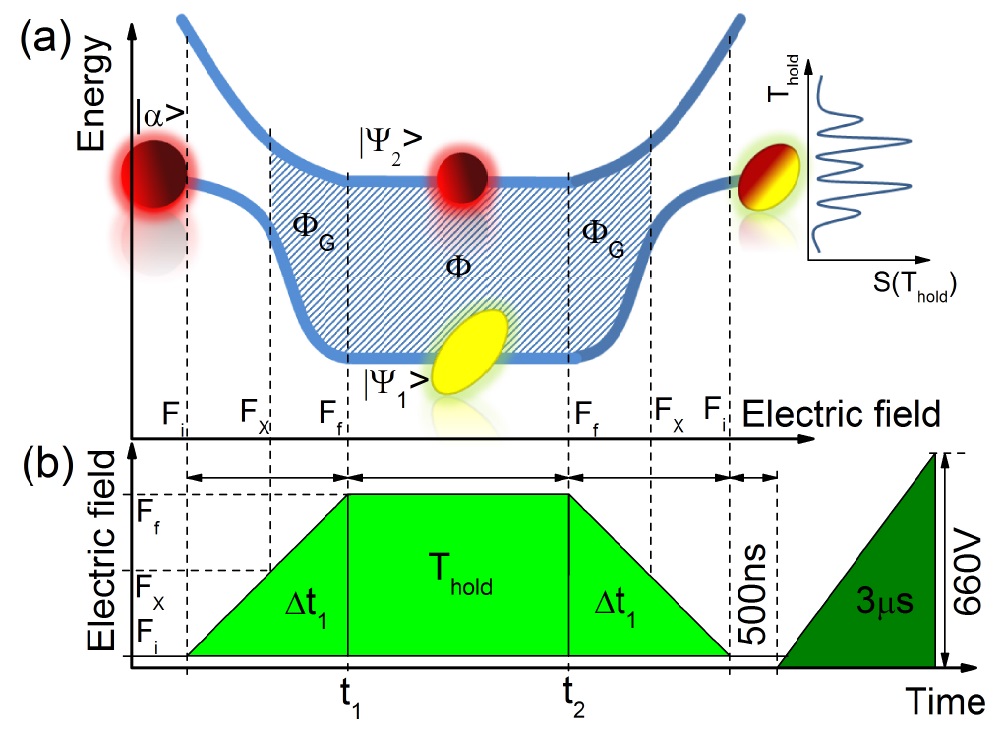} \newline
\vspace{-2em}
\caption{(color online) Atom interference of \emph{n}S and high-\emph{l} Stark Rydberg states through an electric-field-induced avoided crossing. (a) After preparing a Rydberg Stark state $|\alpha\rangle$ at an initial field $F_{i}$, the field is ramped to a variable final field value, $F_{f}$. The relevant atomic states anti-cross at a selected avoided crossing centered at field $F_{X}$. During the ramp time through the crossing, $\Delta$t$_{1}$, the atoms undergo mixed diabatic/adiabatic passage into adiabatic states $|\Psi_{1}\rangle$ and $|\Psi_{2}\rangle$, respectively. After a variable hold time $T_{\rm {hold}} < 60$~ns between times $t_1$ and $t_2$, the electric field is ramped back to $F_{i}$. The interferometric phase $\Phi$, visualized by the hatched area, is measured via the population difference between the two states at the exit of the interferometer. (b) Timing of the ramped electric field (light green) and the state-selective ionization field (dark green) used to read out the interferometer. As explained in the text, a waiting time of 500~ns before the rise of the ionization field (displayed time interval not to scale) enables state-selective detection.}
\end{figure}

Interferometric methods also extend to Rydberg atoms. These highly excited atoms are attractive in field metrology due to their strong response to applied electric fields (polarizabilities typically scale $\propto n^7$\cite{tgallagher}) and microwave/THz fields.
Couplings between Floquet states of thermal Rydberg atoms in microwave fields give rise to St\"uckelberg oscillations~\cite{Stueckelberg} and interference effects in microwave multiphoton excitation~\cite{Baruch,Yoakum}. St\"uckelberg oscillations based on avoided crossings between two-atom energy levels shifted by the dipole-dipole interaction between Rydberg atoms have been investigated in an oscillating radio-frequency field~\cite{van}. Ramsey interferometry involving optical and external-electric-field pulses has been employed to detect Stark-tuned F\"orster resonances and the interaction-induced phase shift of cold rubidium Rydberg atoms ~\cite{jnipper,nipper}. Stark-induced \emph{l}-mixing interferences based on avoided crossings between low-\emph{l} states and nearby Stark manifolds of cold cesium Rydberg atoms have been observed by applying electric-field pulses~\cite{hao}.

In this work we develop a Rydberg-atom interferometer in which avoided crossings in the Stark map are utilized as beam splitters and recombiners, resulting in an interferometer that mimics an optical Mach-Zehnder interferometer.
The coherent state mixing occurs during multiple passages of the atoms through avoided crossings. A time-dependent electric field acts as a control parameter for the passage behavior in the avoided crossings. In the second passage, the accumulated phase of the wavefunction coherence is mapped into a measurable population difference between the diabatic Rydberg states of the system.
Using the sequence displayed in Fig.~1, we study the dependence of the interferometric response on the electric field inside the interferometer loop, $F_f$, the ramp time through the avoided crossings, $\Delta t_1$, and the electric-field hold time $T_{\rm {hold}}$ between the two field ramps. The coherence frequencies, obtained by Fast Fourier Transforms (FFTs) of the interferometric signals, are compared with the results of a theoretical model. We exhibit the ability of the interferometer to measure energies of ``hidden'' levels that cannot be directly optically excited as well as other applications.

The cesium atoms are trapped in a standard magneto-optical trap (MOT); for details see Ref.~\cite{hao}. As shown in Fig.~2, the Rydberg atoms are initially prepared in the well-defined adiabatic state $| \alpha \rangle$, in the initial electric field $F_{i}$. When the electric field is ramped to its final value, $F_{f}$, during the ramp time $\Delta t_{1}$, mixed diabatic/adiabatic passage through the selected avoided crossing coherently splits the wavefunction into the adiabatic states $|\Psi_{2}\rangle$, which predominantly has 49S$_{1/2}$-character, and $|\Psi_{1}\rangle$, which is a hydrogen-like linear Stark state with predominantly high-\emph{l} character.
\begin{figure}[htb]
\vspace{-1em}
\centering
\includegraphics[width=0.5\textwidth]{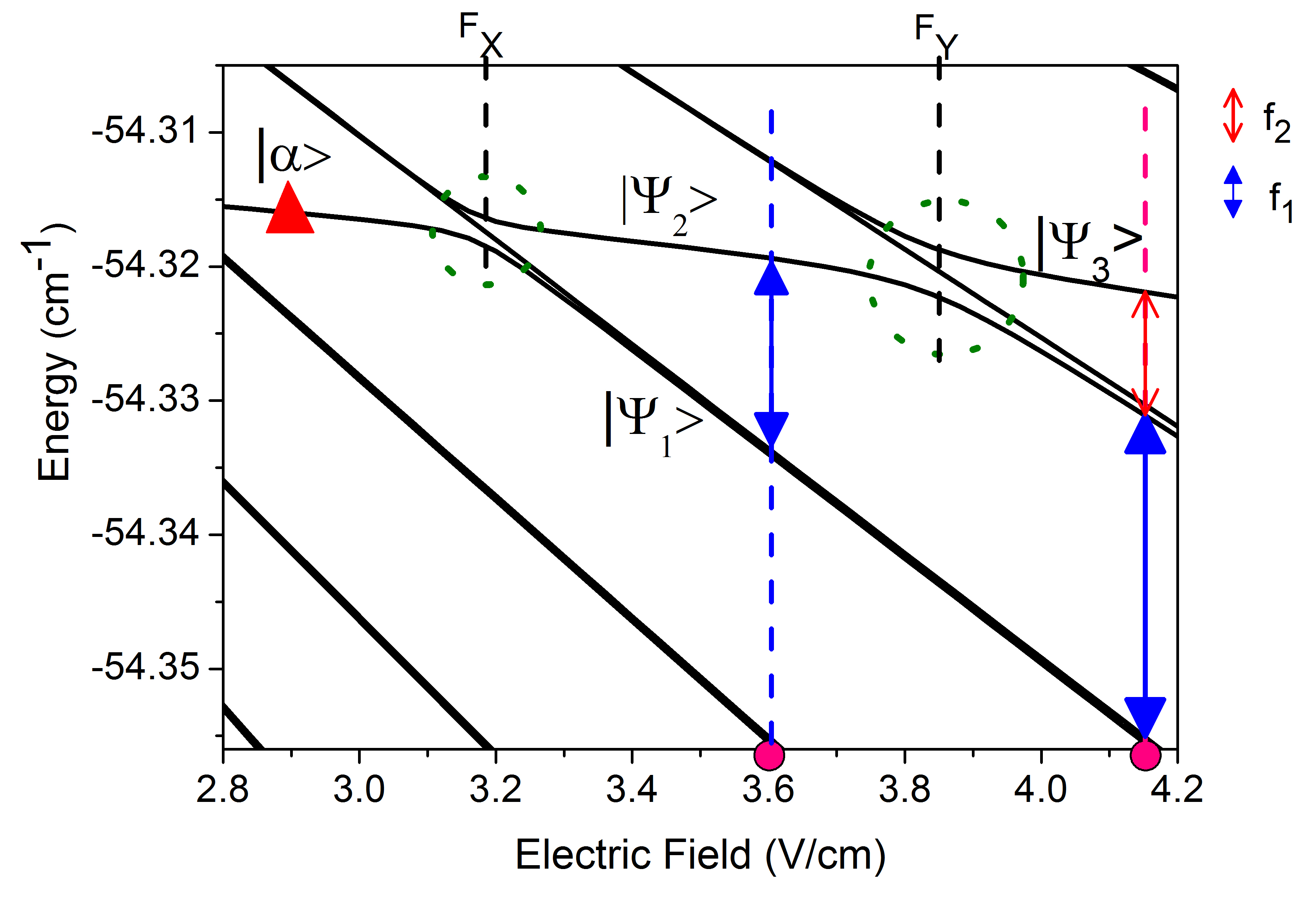} \newline
\vspace{-2em}
\centering
\caption{(color online) Calculated Stark map in the vicinity of the 49S$_{1/2}$ state and the n=45 manifold of cesium over the field range 2.8-4.2~V/cm. The range covers two avoided crossings centered at $F_{X}$ = 3.19~V/cm and $F_{Y}$ = 3.85~V/cm, respectively. Atoms are initially excited into state
$\vert \alpha \rangle$ at electric field $F_{i}$ = 2.9~V/cm. We study the regions $F_{X} < F_{f} < F_{Y}$ and $F_{f} > F_{Y}$ (corresponding to the dots on the field axis). In both regions the interferometric signal
contains the coherence frequency f$_{1}$ (corresponding to the blue wide arrow), which originates in coherent splitting and recombination at field $F_{X}$. The interferometric signal in the second region exhibits additional frequencies
f$_{2}$ (red thin arrow) and f$_{1}$+f$_{2}$, which arise due to additional splitting and recombination at field $F_{Y}$.}
\end{figure}

 The selected avoided crossing is centered at $F_X = 3.19~$V/cm (dotted circle in Fig.~2). After the holding time $T_{\rm {hold}}$ at $F_{f}$, the electric field is switched back to $F_{i}$, using the same ramp time $\Delta t_{1}$. The atoms pass the selected avoided crossing twice, namely at times near $t_1$ and $t_2$. The differential phase $\Phi$ between $|\Psi_{1}\rangle$ and $|\Psi_{2}\rangle$ accumulated between $t_1$ and $t_2$, indicated by the hatched area in Fig.1~(a), is given by
\begin{equation}\label{eq:1}
   \Phi(T_{\rm {hold}}) = \Phi_{G}+\frac{1}{\hbar} \int_{t_1}^{t_2=t_1+T_{\rm {hold}}} (E_2-E_1) dt \quad,
\end{equation}
where $\Phi_{G}$ includes a geometrical phase that depends on the passage through the crossing. The integral in Eq.~(1) reflects the dynamical phase accumulated during the hold time. The energy difference $E_2-E_1$ between the adiabatic states $|\Psi_{2}\rangle$ and $|\Psi_{1}\rangle$ at the final field is approximately given by
 \begin{equation}\label{eq:2}
  E_2-E_1 = - \Delta d (F_f - F_X) \quad,
 \end{equation}
where $\Delta d$ is the difference between the electric-dipole moments of the adiabatic states $|\Psi_{2}\rangle$ and $|\Psi_{1}\rangle$ at field $F_f$. It is seen that for fixed $\Delta t_{1}$ the interferometric phase
\begin{equation} \label{eq:3}
   \Phi(T_{\rm {hold}}) \approx \Phi_{G} - \frac{1}{\hbar} \Delta d (F_f - F_X) T_{\rm {hold}} \quad,
\end{equation}
{\sl{i. e.}} it accumulates at a rate largely given by the final-field value $F_f$. Small deviations occur because the electric dipole moments of the adiabatic states are not exactly fixed. Each time the electric field passes through $F_{X}$, the field at which the selected crossing is centered, the  adiabatic states $|\Psi_{1}\rangle$ and $|\Psi_{2}\rangle$ are subject to coherent mixing. The mixing is analogous to beam splitting and recombination in optical interferometers. At the output of the interferometer, we detect the fraction of the atom population that exits in the original S-like state $\vert \alpha \rangle$. Experimentally the 500~ns waiting period before detection is critical because it enables state-selective readout of the interferometer by separating the elongated high-\emph{l} Stark state and state $\vert \alpha \rangle$ through $m$-mixing collisions \cite{Limei}.

In preparation for our experiment, we numerically obtain the Stark energy level structure for cesium and simulate the time evolution
of the wavefunction. For an electric field pointing along $z$ with amplitude $F(t)$ and effective mass $\mu$, the Hamiltonian
\begin{equation}\label{eq:4}
  \hat{H} = \frac{\hat{\bf p}^{2}}{2 \mu} - \frac{1}{\hat{r}} + V_c(\hat{r}) + V_{FS} +
  F (t) \hat{z} \quad ,
\end{equation}
where $V_c$ is a short-range core potential, $V_{FS}$ is the fine structure, and $F (t) \hat{z}$ is the perturbation due to the time-dependent electric field. In the model we use quantum defects from~\cite{tgallagher}. Figure~2 shows the cesium Stark structure in the vicinity of the $n = 45$ hydrogen-like manifold \cite{mlzimmerman} zoomed into the electric-field range of 2.8 - 4.2~V/cm. The 49S$_{1/2}$-like level encounters two avoided crossings centered at $F_{X}=3.19~$V/cm and $F_{Y}=3.85~$V/cm, which have energy gaps of 58~MHz and 134~MHz, respectively.  The Rydberg atoms are initially prepared in the adiabatic state $\vert \alpha \rangle$ at an electric field $F_{i}$ = 2.9~V/cm. Using this state
as initial wavefunction, we numerically solve the time-dependent Schr\"odinger equation for the sequence in Fig.~1 and obtain the final state after completion of the second ramp, $\vert \Psi_{\rm{ end}} \rangle$. The interferometric signal $S = \vert \langle \alpha \vert \Psi_{\rm{ end}} \rangle \vert^2$ is computed as function of $T_{\rm{hold}}$ for a selection of values for the ramp time  $\Delta t_{1}$ and the field $F_{f}$.

In Fig.~3~(a) we set $F_{f}$ at the center field $F_{X} = 3.19~$V/cm and show the signal $S(T_{\rm {hold}})$ for a range of ramp times $\Delta t_{1}$. The state cutting through the center of the avoided crossing can be largely ignored, as found in Ref.~\cite{Limei}. The signal $S(T_{\rm{hold}})$ oscillates with a period of 17~ns, corresponding to the gap of 58~MHz between the levels at the first avoided crossing. The visibility of the oscillation is maximal for $\Delta t_{1} = 0$, because sudden projection of the
initial adiabatic state $\vert \alpha \rangle$ into the center field $F_{X}$ of the crossing yields amplitudes near $1/\sqrt{2}$ for both coupled levels, which are analogous to the case of a 50/50 beam splitter in an optical interferometer; in that case maximum interference contrast occurs. As $\Delta t_1$ increases, the state evolution becomes increasingly adiabatic and the splitting ratio continuously changes from 50\% adiabatic to 100\% adiabatic. Hence, as seen in Fig.~3~(a), the visibility of the oscillation in $S$ diminishes with increasing $\Delta t_1$.

In Fig.~3~(b) the field $F_{f}=3.51$~V/cm, which is in-between two avoided crossings. As before, the oscillation frequency of $S$ is given by the energy difference between the adiabatic states $|\Psi_{2}\rangle$ and $|\Psi_{1}\rangle$, but the visibility of the oscillation in the signal $S$ peaks at $\Delta t_1 = 16$~ns. In Fig.~3~(b), the interference signal peaks under conditions when the Landau-Zener passage dynamics through the crossing leads to 50$\%$ population in each of the adiabatic states
$|\Psi_{2}\rangle$ and $|\Psi_{1}\rangle$. A straightforward calculation for a two-level Landau-Zener crossing with gaps and slopes as in Fig.~2 shows that parity between diabatic and adiabatic passage probability is indeed expected at $\Delta t_1 = 16$~ns.

The phase shift of the signal reflects a variation of $\Phi_{G}$ in Eq.~\ref{eq:3}. The phase shift does not affect the frequency and the visibility of the interference; in the present work we are not concerned with it.

\begin{figure}[htb]
\vspace{-0.1em}
\centering
\includegraphics[width=0.5\textwidth]{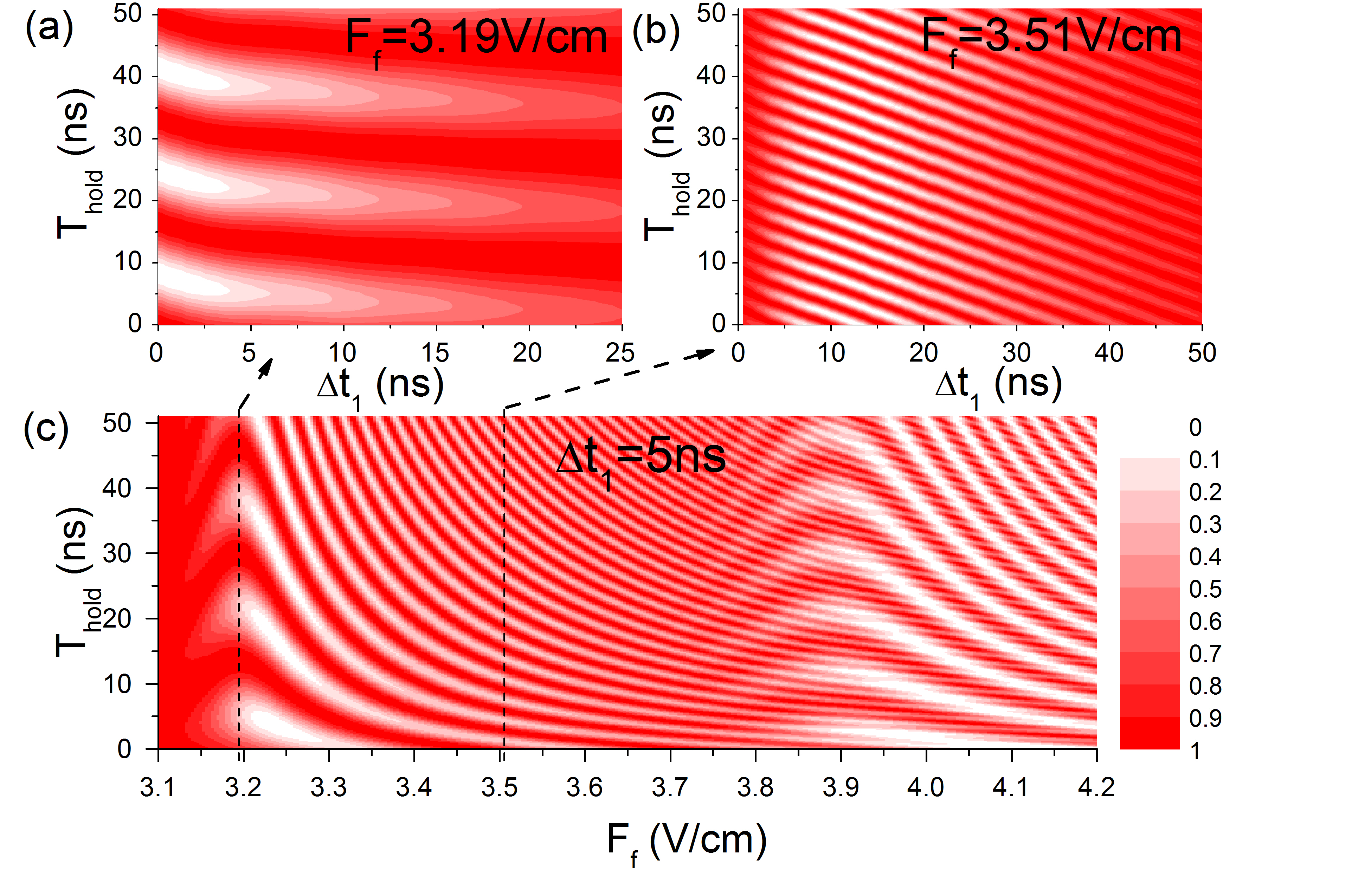} \newline
\vspace{-2em}
\caption{(color online) Simulated atom interference data for different final electric-field values, $F_{\rm f}$. (a) Probability of the system returning to the initial state  $\vert \alpha \rangle $ as a function of ramp duration $\Delta t_{1}$ and holding time T$_{\rm {hold}}$ when ramping the electric field from $F_{i}$=2.9~V/cm to $F_{f} = F_{X} = 3.19$~V/cm (the center field of the first avoided crossing in Fig.~2). (b) Same as (a), but for $F_{f}=3.51$~V/cm (between the two crossings in Fig.~2). (c) Probability of the system returning to the initial state  $\vert \alpha \rangle $ as a function of T$_{\rm{hold}}$ and over a range of $F_{f}$-values, for fixed $F_{i}=2.9$~V/cm and $\Delta t_{1}=5$~ns.}
\end{figure}

From the simulations in Fig.~3~(a), (b) and similar ones not shown we conclude that for our experimental studies a choice of $\Delta t_1 = 5$~ns should yield high interference contrast in $S(T_{\rm{hold}})$ for all values of $F_{f}$. In Fig.~3~(c) we plot $S(T_{\rm{hold}})$ as a function of the final electric field $F_{f}$ for  a fixed $\Delta t_{1}$ = 5~ns. It is seen that interferometric  oscillations in $S$ are clearly visible over a wide range $F_{f}$.

The oscillation frequencies depend on $F_{f}$ in a manner that reflects the  energy splittings in the Stark map. The oscillation frequency increases with $F_{f}$ in the domain $F_{X} \lesssim F_{f} \lesssim F_{Y}$; this frequency corresponds to f$_{1}$ (blue wide arrow) in Fig.~2. When the final field is increased beyond $F_{Y}$, the signal $S(T_{\rm{hold}})$ displays several frequencies. Noting that mixed diabatic/adiabatic passage from $F_{i} = 2.9$~V/cm to $F_{f} \gtrsim F_{Y}$ will generate amplitudes in all three adiabatic states $|\Psi_{1}\rangle$ to $|\Psi_{3}\rangle$ in Fig.~2, we expect to find three frequencies in that domain 
  f$_{1}$, f$_{2}$ (red thin arrow), and their sum $f_1+f_2$. To find the frequencies, in the following analysis of experimental and simulation data we perform FFTs of the signal $S(T_{\rm{hold}})$.

\begin{figure}[bth]
\vspace{-1em}
\centering
\includegraphics[width=0.5\textwidth]{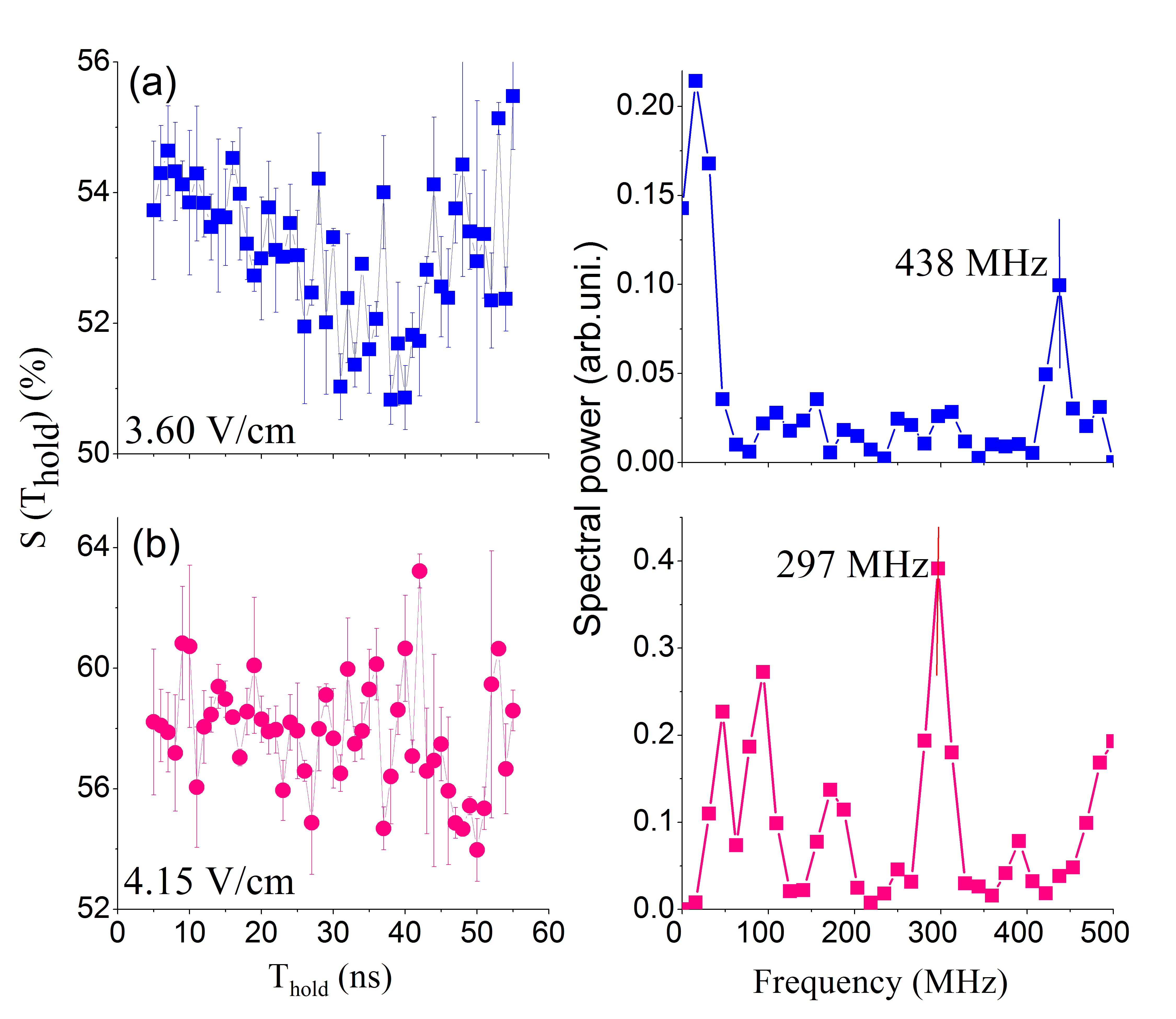} \newline
\vspace{-2em}
\caption{(Color online)  Measured probability of returning to the initially excited S-like Stark state, $S = \vert \langle \alpha \vert \Psi_{\rm{ end}} \rangle \vert^2$, as a function of $T_{\rm { hold}}$ (left panels) and powers of the corresponding FFTs (right panels) for two values of the final electric field, $F_{f}$ = 3.60~V/cm (a) and 4.15~V/cm (b), respectively.}
\end{figure}

In Fig.~4 we show experimental interferometric signals $S(T_{\rm{hold}})$ for $F_{f}$ = 3.60~V/cm [panel (a), left plot] and $F_{f}$ = 4.15~V/cm [panel (b), left plot]. The chosen $F_{i}$ is 2.9~V/cm and $\Delta t_{1}$ equals 5~ns. The plots on the right show the respective spectral power of the FFTs of the measured $S(T_{\rm{hold}})$. To suppress artifacts at low frequencies and spectral side lobes, we subtract the time-averaged values of $S$ and multiply with a standard window function (the Hanning window) before computing the FFTs. Since the experimental data were sampled in 1~ns steps (the smallest step size of the waveform generator used to generate the time-dependent electric field), the experimental data are limited to the range $f < 500$~MHz, the Nyquist frequency.

The peaks in Fig.~4 correspond to frequencies of the interference signal of 438~MHz for 3.60~V/cm and 297~MHz for 4.15~V/cm.
These experimentally observed frequencies agree very well with the Stark-map frequencies labeled $f_1$ and $f_2$ in Fig.~2, respectively. The peak near 20~MHz in Fig.~4~(a) is discarded because it appears to reflect an overall, slow signal drift that occurred while taking the curve $S(T_{\rm{hold}})$.

In Fig.~4~(a) we generate a coherence via mixed diabatic/adiabatic passage through the first crossing in Fig.~2, and the coherence evolves at a frequency given by the Stark splitting between adiabatic states $|\Psi_{2}\rangle$ and $|\Psi_{1}\rangle$ at $F_f=$3.60~V/cm. Our interferometric measurement scheme employed here allows one to measure frequency splittings of states that are optically not excitable due to selection rules. For example, in cesium the linear Stark states (such as $|\Psi_{1}\rangle$) have weak oscillator strength with low-lying $S$ and $P$-levels. The adiabatic passage through the avoided crossing into the optically non-excitable state serves as a way to circumvent optical selection rules: it allows us to probe states that are hidden in optical excitation spectra.

In Fig.~4~(b) the ramp speed is faster due to larger $F_{f}$, and the passage through the first crossing is mostly diabatic. The passage through the second crossing ($F_Y$ in Fig.~2) leads to an adiabatic/diabatic splitting ratio near 50/50, which results in a signal in which the frequency component $f_2$ between levels $|\Psi_{3}\rangle$ and $|\Psi_{2}\rangle$ has a high amplitude ({\sl{i.e.}} the corresponding
oscillation in the interferometric signal $S(T_{\rm{hold}})$ has a high visibility). Fig.~4~(b) therefore shows that our interferometric method offers flexibility in measuring energy-level differences involving a variety of hidden states of interest. This is done by selecting specific avoided crossings and ramp speeds that result in significant populations in the optically unaccessible states of interest.

\begin{figure}[bth]
\vspace{-2em}
\centering
\includegraphics[width=0.5\textwidth]{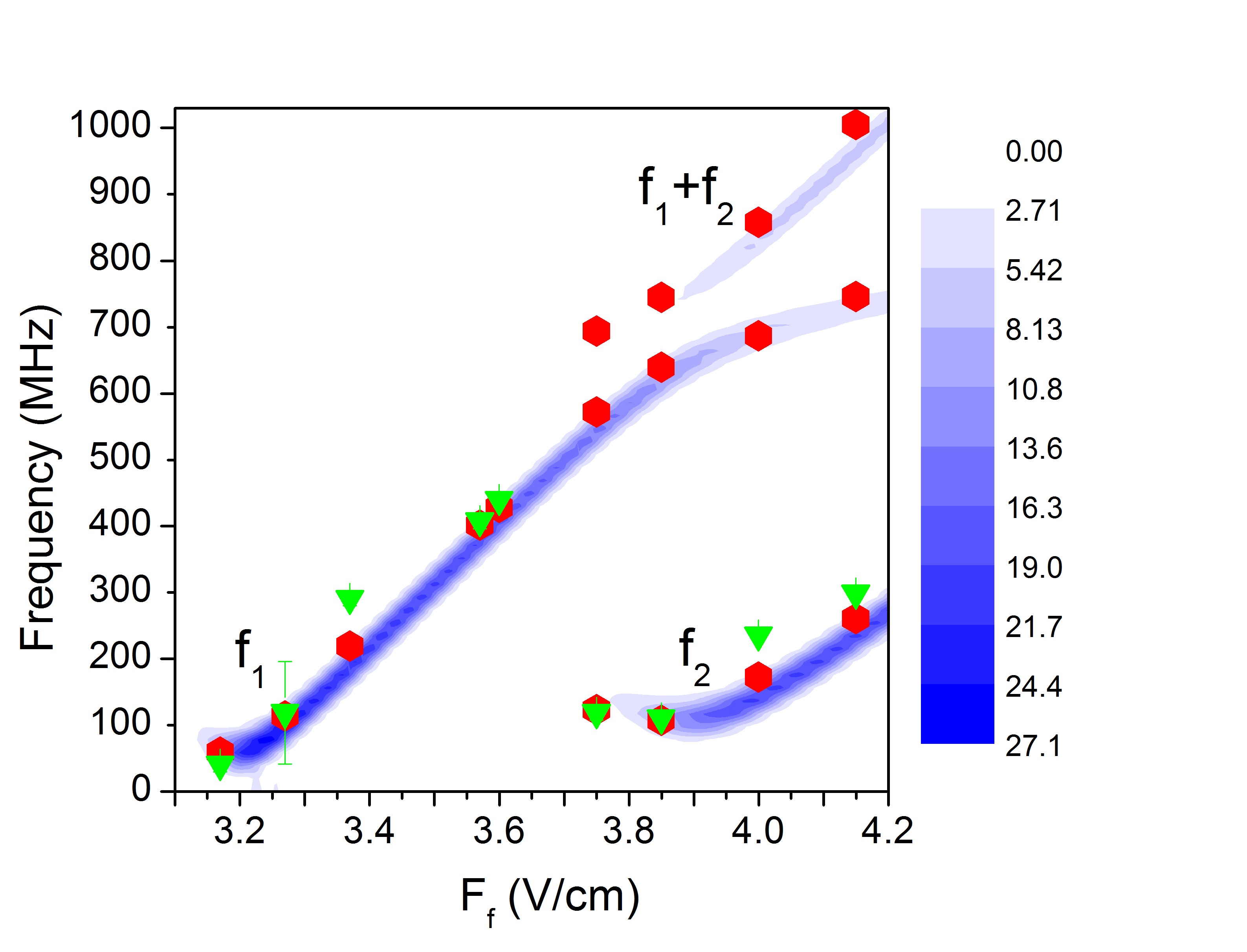} \newline
\vspace{-1em}
\caption{(Color online) Measured coherence frequencies (green triangles) and corresponding calculated frequency differences $f_1$, $f_2$, and $f_1+f_2$ between Stark states (red diamonds) vs. the final electric field, $F_{f}$. The gray-scale plot in the background shows the FFT spectral density obtained from Fig.~3~(c).}
\end{figure}

We have performed a series of additional measurements at different final field $F_{f}$. In Fig.~5 we present the frequency values observed in the FFTs of interferometric signals $S(T_{\rm{hold}})$ measured at different final electric-field values (green triangles). 
Potentially possible aliasing signals in the experiment were not observed. The experimental frequencies are compared with corresponding calculated frequency spacings $f_{1}$, $f_{2}$, and $f_{1}+f_{2}$ (red diamonds). The backdrop shows the FFT of the simulated signals from Fig.~3~(c). The simulated FFTs show signals up to 1~GHz (the simulation has a sampling step size of 0.2~ns and a Nyquist frequency of 2.5~GHz; between 1~GHz and 2.5~GHz the simulated FFTs do not show significant signals). All three types of data in Fig.~5 are consistent with each other. It is noted that only f$_{1}$, f$_{2}$ and $f_{1}+f_{2}$ significantly contribute to the FFTs of the interferometric signal, despite the fact that hundreds of Stark states near the selected avoided crossings are included in the calculation. According to the interpretation given above, this is due to the fact that only the adiabatic levels $|\Psi_{1}\rangle$, $|\Psi_{2}\rangle$ and $|\Psi_{3}\rangle$ become populated due to coupling at the selected avoided crossings.

In conclusion, we have observed the coherence between quantum states using an interferometric method, in which
an external electric field is ramped twice through selected avoided crossings. The interferometric signal
is observed by varying the hold time between the field ramps, in analogy with varying the path-length difference in an optical interferometer. The coherence frequencies observed in the Fourier transforms of the signal reflect the energy-level differences in the underlying Stark map. The method allows us to map out levels that, due to selection rules, are hidden in optical excitation spectra.
Future applications of the atom-interferometric scheme could include metrology of static and radio-frequency (RF) electric fields. To measure RF fields, one can employ state-selective AC shifts (for instance, a state-selective AC shift of the S-like level due to a near-resonance of the RF field with an S to P transition). Conversely, a well-characterized field switch could be used to switch the interferometric phase. Single-atom gates operating along these lines could be combined with interferometers involving Rydberg-atom pairs, whose interactions give rise to level-specific interferometric phases.

The work was supported by the 973 Program (2012CB921603), the NSFC Project for Excellent Research Team (61121064), the NNSF of China (11274209, 61475090, 61378039 and 61378013) and the NSF (PHY-1205559).


\begin{thebibliography} {199}

\bibitem{angelis}M. de Angelis, A. Bertoldi, L. Cacciapuoti, A. Giorgini, G. Lamporesi, M. Prevedelli, G. Saccorotti, F. Sorrentino, and G M Tino, Meas. Sci. Technol. \textbf{20}, 022001 (2009).
\bibitem{hughes}K. J. Hughes, J. H. T. Burke, and C. A. Sackett, Phys. Rev. Lett. \textbf{102}, 150403 (2009).
\bibitem{Geiger}R. Geiger, V. M\'enoret, G. Stern, N. Zahzam, P. Cheinet, B. Battelier, A. Villing, F. Moron, M. Lours, Y. Bidel, A. Bresson, A. Landragin, and P. Bouyer, Nat. Commun. \textbf{2}, 474 (2011).
\bibitem{McGruirk}J. M. McGuirk, G. T. Foster, J. B. Fixler, M. J. Snadden, and M. A. Kasevich, Phys. Rev. A \textbf{65}, 033608 (2002).
\bibitem{Kohel}N. Yu, J. M. Kohel, J. R. Kellogg, and L. Maleki, Appl. Phys. B \textbf{84}, 647 (2006).
\bibitem{Gustavson} T. L. Gustavson, A. Landragin, and M. A. Kasevich, Class. Quantum Grav. \textbf{17}, 2385 (2000).
\bibitem{Lenef1997} A. Lenef, T.D. Hammond, E.T. Smith, M.S.Chapman,  R.A. Rubenstein, D.E.Pritchard, Phys. Rev. Lett. {\textbf 78}, 760 (1997).
\bibitem{Fagaly}R L Fagaly, Rev. Sci. Instrum. \textbf{77}, 101101 (2006).
\bibitem{rkVarma}R. K. Varma, S. B. Banerjee, and A. Ambastha, Eur. Phys. J. D \textbf{66}, 38(2012).
\bibitem{RKV}R. K. Varma, Phys. Rev. E, \textbf{64}, 036608 (2001).
\bibitem{RK}R. K. Varma, A. M. Punithavelu, and S. B. Banerjee, Phys. Rev. E \textbf{65}, 026503 (2002).
\bibitem{Ramsey} N. F. Ramsey, Rev. Mod. Phys. \textbf{62}, 541 (1990).
\bibitem{ADCronin}A. D. Cronin, J. Schmiedmayer and D. E. Pritchard, Rev. Mod. Phys. \textbf{81}, 1051 (2009).
\bibitem{KHornberger}K. Hornberger, S. Gerlich, P. Haslinger, S. Nimmrichter, and M. Arndt Rev. Mod. Phys. \textbf{84}, 157 (2012).
\bibitem{tgallagher} T. Gallagher, \textbf{Rydberg atoms}, Cambridge University Press, Cambridge, U.K. (1994).
\bibitem{Stueckelberg} E. C. G. St\"uckelberg, Helv. Phys. Acta \textbf{5}, 369 (1932).
\bibitem{Baruch} M. C. Baruch and T. F. Gallagher, Phys. Rev. Lett. \textbf{68}, 3515 (1992).
\bibitem{Yoakum} S. Yoakum, L. Sirko and P. M. Koch, Phys. Rev. Lett. \textbf{69}, 1919 (1992).
\bibitem{van} C. S. E. van Ditzhuijzen, Atreju Tauschinsky, and H. B. van Linden van den Heuvell, Phys. Rev. A \textbf{80}, 063407 (2009).
\bibitem{jnipper} J. Nipper, J. B. Balewski, A. T. Krupp, B. Butscher, R. L\"ow, and T. Pfau, Phys. Rev. Lett. \textbf{108}, 113001 (2012).
\bibitem{nipper} J. Nipper, J. B. Balewski, A. T. Krupp, S. Hofferberth, R. Lo¡§w, and T. Pfau, Phys. Rev. X \textbf{2}, 031011 (2012).
\bibitem{hao} H. Zhang, L. Wang, L. Zhang, C. Li, L. Xiao, J. Zhao, and S. Jia,  Phys. Rev. A \textbf{87}, 033405 (2013).
\bibitem{Limei}L. Wang, H. Zhang, L. Zhang, C. Li, Y. Yang, J. Zhao, G. Raithel, and S. Jia, arXiv:1504.00518v1.
\bibitem{mlzimmerman} M. L. Zimmerman, M. G. Littman, M. M. Kash, and D. Kleppner, Phys. Rev. A \textbf{20}, 2251 (1979).
\end{thebibliography}
\end{document}